\documentclass[
    journal,
    twocolumn,
    ]{IEEEtran}
\usepackage{graphicx}
\usepackage{amsmath,amssymb,amsfonts}
\usepackage{epsfig}
\usepackage{epstopdf}
\usepackage{amsthm}
\usepackage{afterpage}
\usepackage{verbatim}
\usepackage{psfrag}
\usepackage{color}
\usepackage{xcolor}
\usepackage{cite}
\usepackage{orcidlink}
\usepackage{setspace}
\usepackage{adjustbox}
\usepackage{footmisc}
\usepackage{fmtcount}
\usepackage{stackrel}
\usepackage{float}
\usepackage{breqn}
\usepackage{hyperref}
\usepackage{enumerate}
\usepackage{subcaption}
\usepackage{algorithm}
\usepackage{algpseudocode}
\usepackage{soul}
\usepackage{xcolor}
\usepackage{multirow}
\usepackage[T1]{fontenc}
\usepackage{tabularx,booktabs}

\begin{document}

\title{AI-Assisted ISAC Localization-as-a-Service \\ for 6G UAV-IoT Networks}

\author{Ruhul Amin Khalil\orcidlink{0000-0003-4039-9901}, \IEEEmembership{Senior Member, IEEE} 
\thanks{This work was supported by the Office of the Associate Provost for Research at the United Arab Emirates University (UAEU), UAE. \textit{(Corresponding author: Ruhul Amin Khalil)}}
\thanks{R. A. Khalil is with the Engineering Requirements Unit (ERU), College of Engineering, United Arab Emirates University, Al Ain 15551, UAE. e-mail: ruhulamin@uaeu.ac.ae}

\thanks{Manuscript received ~X, X~X; revised ~X, X~X.}}


\maketitle

\begin{abstract}
Integrated sensing and communication (ISAC) can enable sixth-generation (6G) unmanned aerial vehicle-assisted Internet of Things (UAV-IoT) networks to provide reliable Localization-as-a-Service (LaaS), but activating all aerial/terrestrial anchors and beams increases pilot overhead, energy use, and beam-training delay. This article proposes artificial intelligence (AI)-assisted ISAC resource selection for LaaS (AIRS-LaaS), an edge-intelligent framework that ranks candidate anchor--beam pairs using line-of-sight (LoS) likelihood, signal-to-interference-plus-noise ratio (SINR), sensing confidence, geometry, mobility risk, and resource cost. A lightweight selector then activates only a compact subset before localization. Simulations compare AIRS-LaaS with all-anchor, Fisher information matrix/Cramér--Rao lower bound (FIM/CRLB)-greedy, strongest-SINR, nearest-anchor, and random schemes under LoS/non-line-of-sight (NLoS) conditions and UAV mobility. Results show a balanced localization--communication--overhead tradeoff, while the discussion highlights standard-driven key performance indicators (KPIs), ISAC reporting, localization confidence, fallback operation, AI model management, and privacy-aware data exchange.
\end{abstract}


\section{Introduction}
Sixth-generation (6G) networks are expected to evolve from connectivity-centric systems into intelligent service platforms that jointly support communication, sensing, localization, and computation. The ITU-R IMT-2030 framework identifies ``artificial intelligence (AI) and communication'' and ``integrated sensing and communication'' (ISAC) as key usage scenarios for future mobile systems \cite{itu2023imt2030}, while recent standardization-oriented studies further emphasize ISAC as a central 6G capability \cite{kaushik2024isac_iot}. In parallel, the Third Generation Partnership Project (3GPP) Release 19 has initiated studies on ISAC channel modeling and higher mid-band spectrum as a bridge toward 6G \cite{3gpp2025rel19,heggo2025isac_channel_etsi}. The European Telecommunications Standards Institute (ETSI) has also released ISAC use-case and deployment studies covering sensing modes, deployment scenarios, and sensing-oriented key performance indicators (KPIs) \cite{etsi2025grisc001}. Moreover, 3GPP AI/machine learning (AI/ML) activities are moving toward standardized support for data collection, model monitoring, activation/deactivation, interoperability, and device behavior \cite{3gpp2025aiml,lin2023aiml_nr,huang2025aiml_lcm}. These developments indicate that localization can become a native network service supported by standardized sensing, signaling, and edge intelligence.

Unmanned aerial vehicle (UAV)-assisted Internet of Things (IoT) networks are a promising environment for such localization services. UAVs can serve as flexible aerial anchors that improve geometry and coverage, enabling rapid deployment in smart cities, industrial sites, precision agriculture, transportation, and emergency scenarios. ISAC-enabled IoT networks are particularly relevant because they can reuse radio resources for sensing and communication \cite{kaushik2024isac_iot}. When integrated with terrestrial base stations, roadside units, or IoT gateways, UAVs provide spatial diversity and line-of-sight (LoS) opportunities that are difficult to guarantee in purely terrestrial deployments. Recent UAV-enabled ISAC studies also show that trajectory design, beamforming, sensing time, and communication quality must be jointly optimized under practical energy and backhaul constraints \cite{khalili2024uav_isac}. \textcolor{black}{Moreover, reliable LaaS outputs can support UAV-IoT digital twins by feeding location, uncertainty, sensing-confidence, and link-quality states into virtual models, complementing recent quality-of-experience (QoE)-aware resource-allocation and semantic-communication digital-twin studies \cite{chen2025generative,zhu2026data}.}

Despite these advantages, dense UAV-IoT deployments introduce a critical resource-selection problem, consistent with prior UAV-ISAC resource-allocation and anchor-activation studies \cite{khalili2024uav_isac,khalil2023anchor_activation}. Activating all available anchors and directional beams may improve localization accuracy, but it also increases pilot overhead, beam-training latency, energy consumption, interference, and reporting load. Energy-efficient anchor activation is important for industrial IoT localization \cite{khalil2023anchor_activation}. Similarly, in smart-city ISAC scenarios, non-line-of-sight (NLoS) propagation and parameter uncertainty can degrade localization unless measurement reliability is explicitly considered. Therefore, a practical 6G Localization-as-a-Service (LaaS) framework must decide not only how to estimate device positions, but also which aerial/terrestrial anchors and beams should participate.

Existing localization and ISAC schemes have made progress in robust estimation, NLoS mitigation, multi-target localization, and adaptive UAV-based localization. For example, multidimensional scaling-based ISAC localization improves multi-target localization robustness \cite{khalil2024isac_mds}. However, many approaches assume that useful measurements are already available or separate localization accuracy from communication reliability and signaling cost. In practical 6G systems, anchor--beam selection should jointly account for LoS probability, communication quality, sensing confidence, geometric diversity, mobility risk, and resource consumption. AI/ML is well suited for this task because it can learn link reliability and measurement usefulness from historical and real-time observations while supporting lightweight edge inference and lifecycle-aware deployment \cite{3gpp2025aiml,huang2025aiml_lcm}.

\textcolor{black}{This article proposes AIRS-LaaS, an AI-assisted ISAC LaaS framework for 6G UAV-IoT networks. Unlike existing UAV-IoT/ISAC studies that mainly optimize UAV trajectory, beamforming, sensing/communication resources, or localization after measurements are available \cite{liu2024uav_isac_iot,khalili2024uav_isac,khalil2024isac_mds}, AIRS-LaaS targets the pre-localization anchor--beam selection problem for network-native LaaS. Its novelty lies in an edge-intelligence layer that jointly scores aerial/terrestrial candidate links using communication reliability, sensing confidence, LoS likelihood, geometric contribution, mobility risk, and resource cost, and activates only a compact subset before localization. Thus, AIRS-LaaS provides communication-aware, sensing-aware, and overhead-aware anchor--beam selection that approaches all-anchor localization accuracy while reducing sensing overhead and maintaining communication reliability.}

The main contributions of this article are as follows:
\begin{itemize}
    \item First, a 3D UAV-IoT ISAC system model is introduced, including aerial anchors, terrestrial anchors, directional beams, IoT targets, and communication-aware sensing links.
    \item Second, a lightweight AI-assisted anchor--beam selection algorithm is developed to balance localization accuracy, pilot overhead, and energy consumption.
    \item \textcolor{black}{Third, the proposed framework is evaluated against all-anchor, FIM/CRLB-greedy, strongest-SINR, nearest-anchor, and random-selection baselines, following representative localization, anchor-activation, and UAV-ISAC comparison settings.
    \item Finally, the implications of standardization are discussed, including ISAC sensing reports, localization confidence indicators, anchor capability signaling, AI model lifecycle management, and privacy-aware measurement exchange.}
\end{itemize}

\section{AI-Assisted ISAC Localization-as-a-Service: Motivation and Requirements}
\subsection{Why Localization-as-a-Service in 6G UAV-IoT Networks?}
Future 6G networks are expected to support connectivity, sensing, positioning, intelligence, and service-aware automation. In UAV-assisted IoT networks, accurate and timely localization is essential for smart transportation, precision agriculture, public safety, industrial monitoring, and emergency response. Unlike conventional positioning, Localization-as-a-Service (LaaS) treats positioning as a network-native capability, where UAVs, terrestrial base stations, roadside units, and IoT gateways cooperatively provide location information on demand.

UAVs are particularly useful for LaaS because they can act as mobile aerial anchors with adjustable altitude, flexible trajectories, and improved LoS opportunities. However, UAV-IoT environments are highly dynamic due to node mobility, blockage, limited UAV energy, and time-varying channels; therefore, an effective 6G LaaS framework must jointly consider localization accuracy, communication reliability, mobility, energy consumption, and signaling overhead. 
\textcolor{black}{AIRS-LaaS can also support UAV-IoT digital twins by feeding location, uncertainty, sensing confidence, and link-quality indicators into the twin state, enabling the twin to evaluate anchor--beam choices before field activation. This complements recent quality-of-experience (QoE)-aware resource-allocation and semantic-communication digital-twin frameworks \cite{chen2025generative,zhu2026data}; however, AIRS-LaaS focuses on localization confidence and anchor--beam selection rather than rendering/QoE optimization or semantic content exchange.}

\subsection{Role of ISAC in Joint Communication and Localization}
ISAC provides a natural foundation for LaaS because the same radio infrastructure, spectrum, and waveform resources can support both data transmission and sensing-related measurements \cite{itu2023imt2030,etsi2025grisc001}. In UAV-IoT networks, ISAC-enabled links can provide communication indicators, such as received signal strength, signal-to-interference-plus-noise ratio (SINR), and beam quality, together with localization-related measurements, such as time of arrival (ToA), angle of arrival (AoA), Doppler shift, and sensing confidence \cite{kaushik2024isac_iot}.

This joint operation is attractive for dense IoT deployments because separate communication and localization procedures may create excessive overhead. By reusing communication pilots, beam-training feedback, and sensing reports, ISAC can reduce duplicated signaling while improving situational awareness. For UAV-assisted networks, ISAC also enables geometry-aware localization, sensing-assisted beam management, and mobility-aware resource allocation, making it suitable for service-oriented 6G networking \cite{liu2024uav_isac_iot,khalili2024uav_isac}.

\subsection{Why AI-Assisted Anchor--Beam Selection is Needed?}
Although ISAC improves the availability of sensing and localization information, activating all candidate anchors and beams is not practical. Dense UAV-IoT networks may include several UAVs, terrestrial anchors, and directional beams for each target device. Using all of them may improve localization accuracy, but it also increases pilot overhead, beam-training latency, energy consumption, interference, and reporting load, creating a fundamental accuracy--overhead tradeoff.

\begin{figure*}[t]
    \centering
    \includegraphics[width=0.95\linewidth]{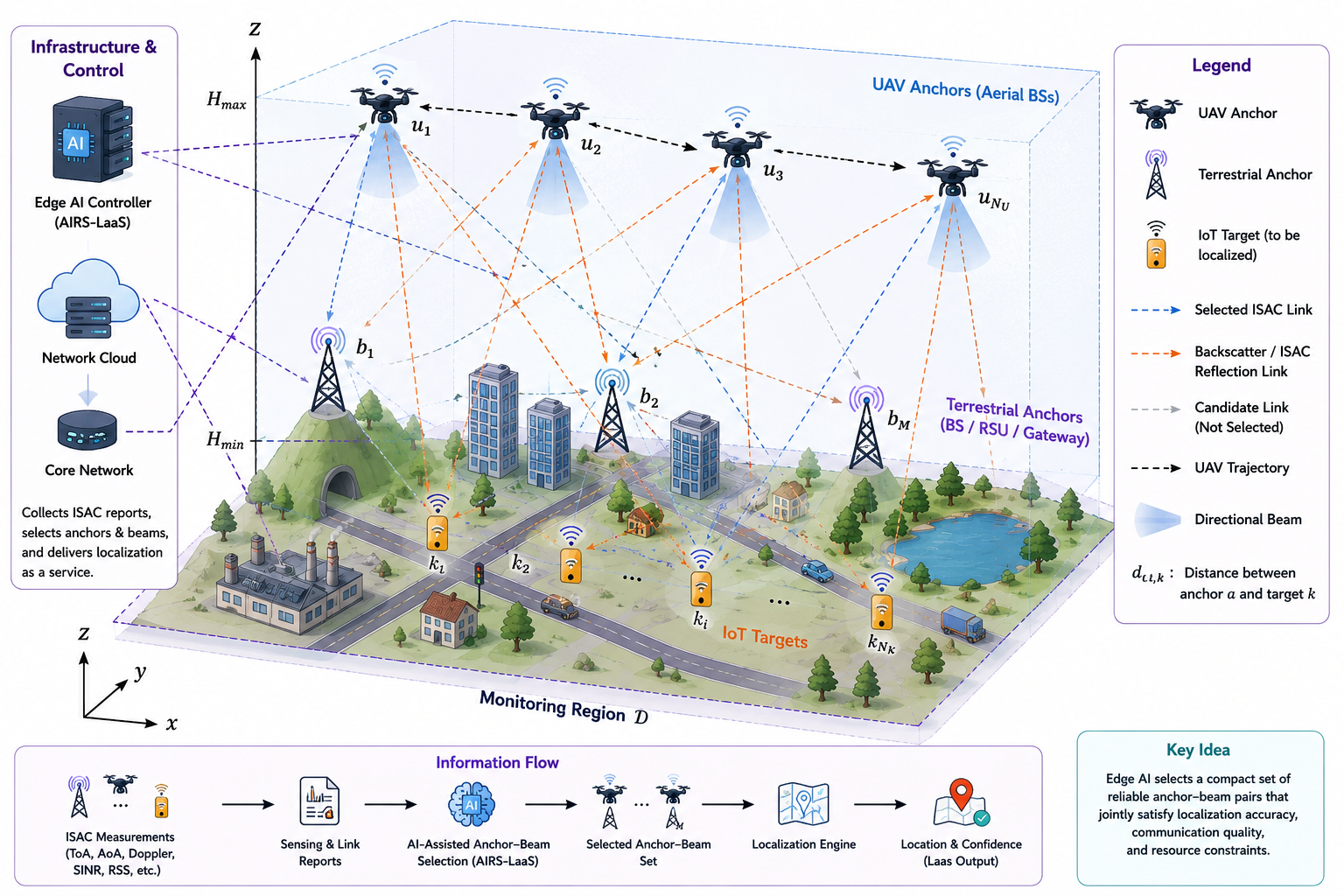}
    \caption{AI-assisted ISAC Localization-as-a-Service system model for 6G UAV-IoT networks, including UAV anchors, terrestrial anchors, IoT devices, directional beams, ISAC measurements, and an edge intelligence layer.}
    \label{fig:system_model}
\end{figure*}

Conventional anchor-selection methods often rely on distance, received signal strength, geometric dilution, or fixed LoS/NLoS assumptions. However, in practical UAV-IoT systems, the best anchor for communication is not always the best anchor for localization. A strong-SINR link may provide poor geometric diversity, while a geometrically useful anchor may suffer from blockage or beam misalignment~\cite{lin2023aiml_nr}. 
\textcolor{black}{AI assistance is therefore used as a lightweight decision layer that learns the joint usefulness of each anchor--beam pair from multiple features, including LoS probability, SINR, sensing confidence, geometry, mobility risk, and resource cost, rather than relying on a single communication or distance metric.}

\textcolor{black}{In the proposed framework, AI does not replace the localization estimator; instead, it selects informative and reliable anchor--beam pairs before localization is performed. This design is suitable for standards-oriented deployment because it operates with measurable network-side indicators and can support model monitoring, fallback, and lifecycle management as envisioned in emerging AI/ML-enabled wireless systems \cite{3gpp2025aiml}.}

\subsection{Design Requirements for Practical Deployment}
A practical AI-assisted ISAC LaaS framework should be communication-aware, overhead-efficient, robust to mobility/NLoS conditions, privacy-aware, and standardization-friendly \cite{kaushik2024isac_iot,etsi2025grisc001}. Selected anchors should provide reliable data links and useful localization measurements under pilot, energy, and latency constraints, using reportable indicators such as SINR, beam quality, sensing confidence, localization uncertainty, anchor capability, and AI model status. 
\textcolor{black}{For standards-driven operation, the edge-intelligence layer can expose compact KPIs, including localization reliability/confidence, LaaS latency, communication outage, active-anchor ratio, sensing-report overhead, energy cost per update, AI inference delay, model confidence, and fallback rate \cite{etsi2025grisc001,huang2025aiml_lcm}.}

\textcolor{black}{In dynamic UAV-IoT deployments, AIRS-LaaS is intended to run mainly at the edge controller, while UAVs report compact sensing and link-quality features instead of executing heavy AI models onboard. This reduces UAV computation, memory, battery, and backhaul burden, but requires robustness to delayed or stale reports \cite{khalili2024uav_isac}. For privacy-aware exchange, anchors should report only necessary features, such as SINR, beam ID, sensing confidence, uncertainty, and pseudonymized target identifiers, while supporting authentication, integrity protection, encryption, access control, and limited edge-side retention. These requirements motivate the proposed anchor--beam selection framework and support LaaS monitoring, capability exposure, and AI model lifecycle management in 3GPP/ETSI-oriented deployments \cite{3gpp2025aiml,huang2025aiml_lcm}.}

\section{System Model}
\subsection{Network Architecture}
We consider a 6G UAV-IoT network in which localization is provided as a network-native service through ISAC, as illustrated in Fig.~\ref{fig:system_model}. The network consists of UAV anchors $\mathcal{U}$, terrestrial anchors $\mathcal{B}$, and IoT target devices $\mathcal{K}$. The complete anchor set is denoted by $\mathcal{A}=\mathcal{U}\cup\mathcal{B}$. UAVs provide flexible aerial coverage and improved geometric diversity, while terrestrial anchors, such as base stations, roadside units, or IoT gateways, provide stable reference points. An edge controller collects sensing and communication reports from candidate anchors and performs AI-assisted anchor--beam selection. \textcolor{black}{The edge controller may be colocated with a base station, roadside unit, or mobile-edge computing node, allowing anchor selection to be performed close to the radio-access network.}

Each anchor $a\in\mathcal{A}$ is equipped with directional beams selected from a beam codebook $\mathcal{M}_a$. For a target device $k\in\mathcal{K}$, each anchor--beam pair $(a,m)$ forms a candidate ISAC link. The position of anchor $a$ is denoted by $\mathbf{p}_a=[x_a,y_a,h_a]^T$, while the unknown position of target device $k$ is denoted by $\mathbf{p}_k=[x_k,y_k,h_k]^T$. Since only a subset of candidate links can be activated under pilot, energy, and latency constraints, the system objective is to select reliable and informative anchor--beam pairs that jointly support localization and communication. \textcolor{black}{This setting captures both aerial flexibility and terrestrial stability, which are essential for UAV-IoT localization under dynamic link and mobility conditions.}

\subsection{ISAC Measurement and Link Model}
Each candidate anchor--beam pair provides both communication- and sensing-related information. Communication indicators include received signal strength, signal-to-interference-plus-noise ratio (SINR), beam quality, and link stability. Sensing indicators include time of arrival (ToA), angle of arrival (AoA), angle of departure (AoD), Doppler shift, and sensing confidence. These measurements can be obtained from communication pilots, beam-training signals, or ISAC sensing reports \cite{kaushik2024isac_iot,liu2024uav_isac_iot}. \textcolor{black}{In practice, these indicators may be reported periodically or triggered by mobility, blockage, beam degradation, or localization service requests.}

For the link between anchor $a$ and target $k$, the measured range is expressed as
\begin{equation}
\hat{d}_{a,k}= \|\mathbf{p}_k-\mathbf{p}_a\| + e_{a,k},
\end{equation}
where $e_{a,k}$ denotes the measurement error caused by thermal noise, multipath, synchronization uncertainty, and non-line-of-sight (NLoS) propagation. Angular measurements are similarly affected by beam resolution, array size, and beam misalignment. Instead of treating all measurements as equally reliable, each candidate link is represented by the feature vector
$\mathbf{x}_{a,m,k} =
[\gamma_{a,m,k}, \hat{d}_{a,k}, \hat{\theta}_{a,k}, q_{a,m,k}, \rho_{a,k}, c_{a,m}]$,
where $\gamma_{a,m,k}$ is the SINR, $\hat{\theta}_{a,k}$ is the estimated angle, $q_{a,m,k}$ is the sensing-confidence indicator, $\rho_{a,k}$ is the predicted LoS probability, and $c_{a,m}$ is the resource cost of activating the corresponding anchor--beam pair. \textcolor{black}{These features are used by the proposed AI-assisted selection algorithm to distinguish reliable links from geometrically useful yet unstable ones.}

\subsection{Communication-Aware Localization Model}
The localization service estimates the position of each target using only the selected anchor--beam pairs. Let $\mathcal{S}_k \subseteq \{(a,m):a\in\mathcal{A},m\in\mathcal{M}_a\}$ denote the selected set for target $k$. The position estimate is obtained from the ISAC measurements collected over $\mathcal{S}_k$, including range, angle, or hybrid range--angle information. Depending on the available measurements, the localization engine may use weighted least squares, hybrid ToA/AoA estimation, or multidimensional scaling-based localization \cite{khalil2024isac_mds}. \textcolor{black}{Measurement weights can be assigned based on sensing confidence and link reliability, so uncertain or NLoS-prone measurements have reduced influence on the final estimate.}

Unlike conventional localization, the selected links must be useful for both positioning and communication. A high-SINR link may not improve localization if it offers limited geometric diversity, whereas a geometrically useful link may be unreliable due to NLoS propagation or beam misalignment. Therefore, selected anchors should meet a minimum communication quality requirement and contribute to localization accuracy. This joint view is important in UAV-IoT networks, where mobility, blockage, and energy constraints make fixed anchor selection inefficient.

The localization error for target $k$ is defined as
\begin{equation}
\varepsilon_k = \|\hat{\mathbf{p}}_k-\mathbf{p}_k\|,
\end{equation}
where $\hat{\mathbf{p}}_k$ is the estimated position. In the simulation section, average and percentile-based localization errors are used to evaluate the reliability of the proposed framework under LoS/NLoS propagation and mobility conditions.

\subsection{Resource and Overhead Model}
Activating more anchors and beams generally improves measurement diversity but increases pilot overhead, beam-training delay, reporting load, and energy consumption. Let $z_{a,m,k}\in\{0,1\}$ indicate whether anchor--beam pair $(a,m)$ is selected for target $k$. The total resource cost for localization-service provisioning is
\begin{equation}
C_k = \sum_{a\in\mathcal{A}}\sum_{m\in\mathcal{M}_a} z_{a,m,k} c_{a,m}.
\end{equation}
The selected set must satisfy a resource budget $C_k \leq C_{\mathrm{max}}$ and a minimum communication-quality condition, such as $\gamma_{a,m,k}\geq \gamma_{\mathrm{min}}$ for selected links. \textcolor{black}{The cost term $c_{a,m}$ can represent pilot use, beam-training effort, sensing-reporting load, or energy expenditure, depending on the deployment objective.}

This model captures the main tradeoff addressed in this article: all-anchor selection may improve localization accuracy but incurs excessive signaling and energy cost, while aggressive anchor reduction may degrade reliability. The proposed AIRS-LaaS algorithm, introduced next, selects a compact subset of anchor--beam pairs that preserves localization accuracy, satisfies communication constraints, and reduces ISAC overhead. \textcolor{black}{The formulation also supports standardization-oriented reporting by relying on practical indicators such as SINR, beam quality, sensing confidence, LoS probability, localization uncertainty, and anchor capability.}

\section{Proposed AIRS-LaaS Algorithm}
This section presents the proposed \emph{AI-Assisted ISAC Resource Selection for Localization-as-a-Service} (AIRS-LaaS) algorithm. The main idea is to avoid activating all available UAV and terrestrial anchors. Instead, the edge-intelligence layer ranks candidate anchor--beam pairs according to their expected contribution to localization accuracy, communication reliability, and resource efficiency. The selected subset is then used by the localization engine to estimate the target position and update link statistics for future decisions.

\subsection{Feature Extraction for Anchor--Beam Pairs}
For each IoT target $k$, the edge controller first constructs a candidate set of anchor--beam pairs from UAV and terrestrial anchors. Each pair $(a,m)$, where $a$ denotes the anchor and $m$ the beam index, is represented by a compact feature vector derived from ISAC sensing reports and communication feedback. The vector includes physical-layer and localization-oriented indicators, such as SINR, received signal strength, beam quality, estimated range, angle information, Doppler shift, sensing confidence, predicted LoS probability, and resource cost.

The feature vector for a candidate pair is written as
$\mathbf{x}_{a,m,k} =
[\gamma_{a,m,k}, q_{a,m,k}, \rho_{a,k}, g_{a,k}, r_{a,m,k}, c_{a,m}],$
where $\gamma_{a,m,k}$ is the communication quality, $q_{a,m,k}$ is the sensing-confidence value, $\rho_{a,k}$ is the predicted LoS probability, $g_{a,k}$ represents the geometric contribution of anchor $a$, $r_{a,m,k}$ denotes the mobility or beam-misalignment risk, and $c_{a,m}$ is the pilot, energy, or beam-training cost. These indicators can be obtained from beam management reports, sensing measurements, localization residuals, and historical link quality observations \cite{3gpp2025aiml,lin2023aiml_nr}.

\subsection{AI-Assisted Link and Sensing-Utility Scoring}
\textcolor{black}{The extracted features are passed to a lightweight AI-assisted scoring block at the edge controller. The reported SINR, sensing confidence, LoS probability, geometric contribution, mobility risk, and resource cost are measured or reported input features; the AI block does not predict these raw values. Instead, it learns/calibrates the mapping from these features to the final utility score $\eta_{a,m,k}$. For the reported results, a lightweight gradient-boosting regressor is used because AIRS-LaaS relies on low-dimensional tabular features. The normalized weighted expression below serves as an interpretable scoring layer, with weights trained/calibrated offline and fixed during online inference.}

The utility score reflects the expected usefulness of each anchor--beam pair for joint localization and communication:
$
\eta_{a,m,k}
=
w_1 \tilde{\gamma}_{a,m,k}
+
w_2 \tilde{q}_{a,m,k}
+
w_3 \tilde{\rho}_{a,k}
+
w_4 \tilde{g}_{a,k}
-
w_5 \tilde{r}_{a,m,k}
-
w_6 \tilde{c}_{a,m},
$
where $\eta_{a,m,k}$ is the utility score, $\tilde{(\cdot)}$ denotes normalized features, and $w_1,\ldots,w_6$ are tunable weights. The positive terms reward reliable communication, high sensing confidence, LoS availability, and useful geometry, while the negative terms penalize mobility risk, beam instability, and resource cost. The weights can be configured according to service requirements; for example, safety-critical localization may prioritize accuracy and reliability, whereas massive IoT monitoring may prioritize energy and overhead reduction.

\subsection{Anchor--Beam Selection Under Resource Constraints}
After computing the utility scores, AIRS-LaaS ranks all candidate anchor--beam pairs and selects a compact subset for localization. The selection is subject to communication and resource constraints, including a minimum SINR, a maximum pilot budget, a maximum number of active beams, and a minimum number of anchors required for reliable localization. This transforms anchor selection into a practical resource-control problem rather than a purely geometric positioning problem.

A low-complexity greedy implementation is suitable for real-time UAV-IoT networks. The edge controller first removes candidate links that do not satisfy the minimum SINR or sensing-confidence requirements. It then sorts the remaining candidates by utility score and progressively activates the best pairs until the pilot or energy budget is reached. To avoid selecting multiple links with redundant geometry, a diversity check is included so that newly selected anchors improve the spatial distribution of measurements. \textcolor{black}{Spatial diversity is quantified by the incremental Fisher-information gain, $\Delta_i=\log\det(\mathbf{J}_{\mathcal{S}_k\cup\{i\}}+\epsilon\mathbf{I})-\log\det(\mathbf{J}_{\mathcal{S}_k}+\epsilon\mathbf{I})$, where $\mathbf{J}_{\mathcal{S}_k}$ is the localization Fisher information matrix of the selected set. A candidate pair is accepted only if $\Delta_i>0$ and the SINR and resource constraints are satisfied, thereby reducing geometric redundancy and dilution of precision.} This is important because high-SINR links alone may not provide accurate localization if the selected anchors are poorly distributed.

\begin{algorithm}[t]
\footnotesize
\caption{\footnotesize{AIRS-LaaS: AI-Assisted ISAC Resource Selection.}}
\label{alg:airs_laas}
\begin{algorithmic}[1]
\Require Candidate anchors $\mathcal{A}$, beam codebooks $\mathcal{M}_a$, target $k$, pilot budget $C_{\mathrm{max}}$, SINR threshold $\gamma_{\mathrm{min}}$
\Ensure Selected anchor--beam set $\mathcal{S}_k$
\State Initialize $\mathcal{S}_k \leftarrow \emptyset$ and $C_k \leftarrow 0$
\For{each anchor--beam pair $(a,m)$}
    \State Collect ISAC and communication features $\mathbf{x}_{a,m,k}$
    \State Estimate LoS probability, sensing confidence, mobility risk, and link quality
    \State Compute utility score $\eta_{a,m,k}$
\EndFor
\State Remove candidates with $\gamma_{a,m,k}<\gamma_{\mathrm{min}}$ or low sensing confidence
\State Sort remaining candidates in descending order of $\eta_{a,m,k}$
\For{each ranked pair $(a,m)$}
    \If{$C_k+c_{a,m}\leq C_{\mathrm{max}}$ and geometry diversity is improved}
        \State Add $(a,m)$ to $\mathcal{S}_k$
        \State Update $C_k \leftarrow C_k+c_{a,m}$
    \EndIf
\EndFor
\State Estimate target position using measurements from $\mathcal{S}_k$
\State Update link statistics using localization residuals and communication feedback
\end{algorithmic}
\end{algorithm}

\textcolor{black}{Let $N_{\mathrm{c}}$ denote the number of candidate anchor--beam pairs and $L$ the maximum selected pairs for a target. AIRS-LaaS requires $O(N_{\mathrm{c}})$ feature scoring, $O(N_{\mathrm{c}}\log N_{\mathrm{c}})$ sorting, and at most $O(LN_{\mathrm{c}})$ diversity/constraint checking, giving per-target complexity $O(N_{\mathrm{c}}\log N_{\mathrm{c}}+LN_{\mathrm{c}})$ and memory $O(N_{\mathrm{c}})$. Since $L\ll N_{\mathrm{c}}$, the algorithm is suitable for edge-side real-time selection.}

\begin{figure*}[!t]
    \centering

    \subfloat[\protect\textcolor{black}{UAV ascent/descent impact on overhead.}]
    {
        \includegraphics[width=0.4\textwidth]
        {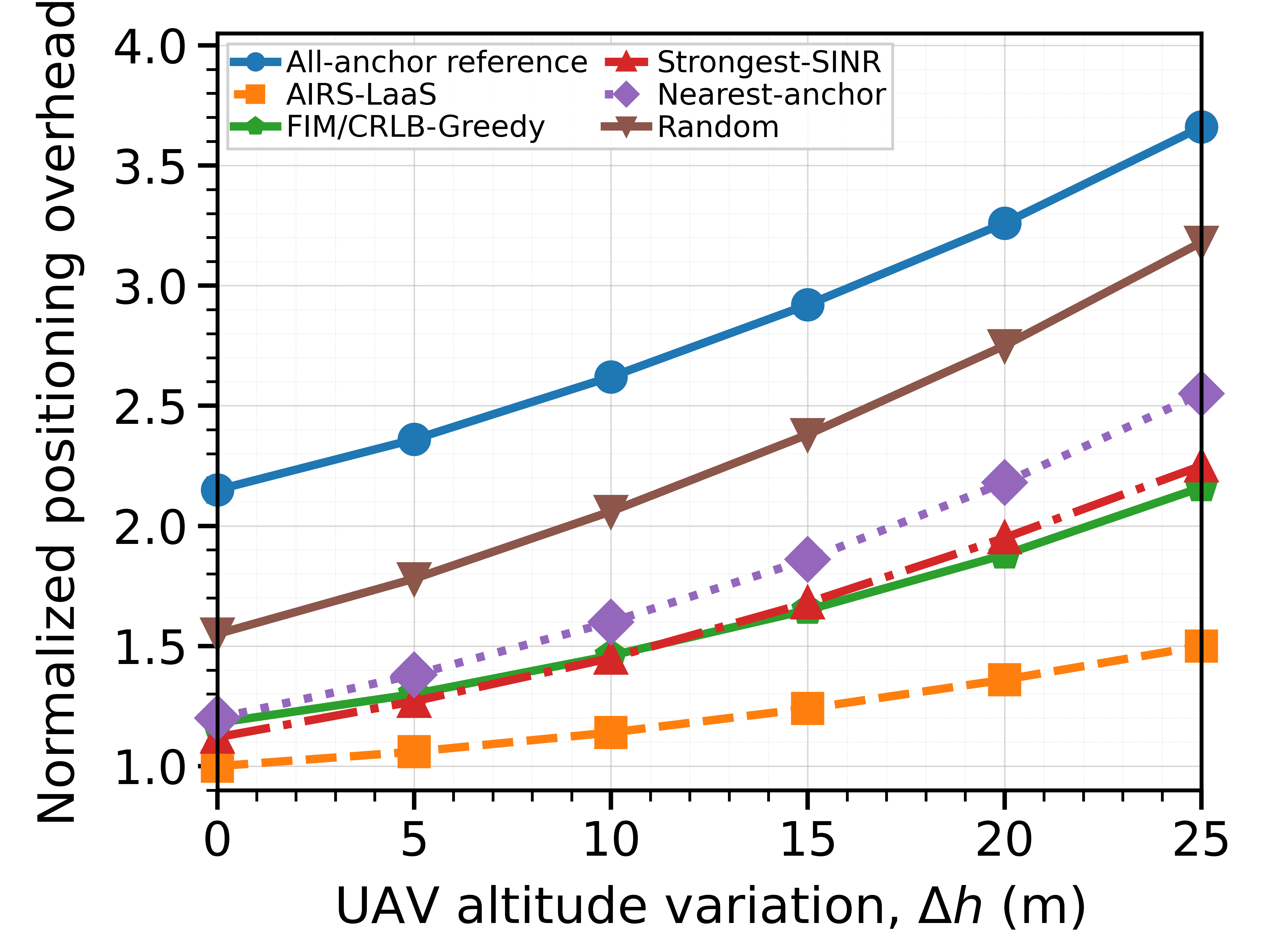}
        \label{fig:dynamic_altitude_overhead}
    }
    \hspace{1.0cm}
    \subfloat[\protect\textcolor{black}{Dense candidate-set impact on outage.}]
    {
        \includegraphics[width=0.4\textwidth]
        {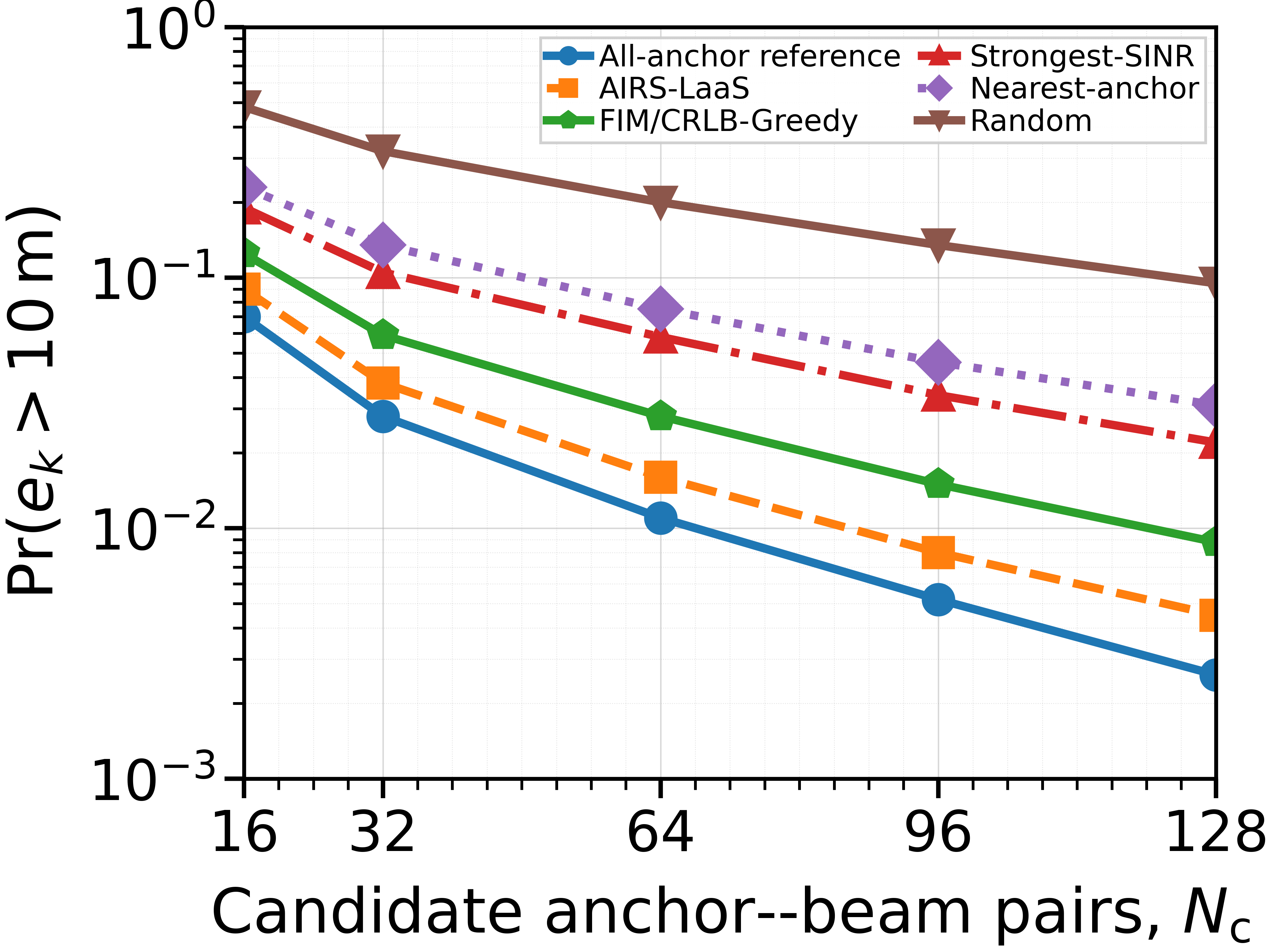}
        \label{fig:dense_candidate_scalability}
    }

    \caption{\textcolor{black}{Robustness evaluation under UAV mobility and dense candidate sets.}}
    \label{fig:mobility_dense_results}
\end{figure*}

\subsection{Localization Update and Feedback Learning}
Once the anchor--beam set $\mathcal{S}_k$ is selected, the localization engine estimates the target position from the available ISAC measurements using weighted least squares, hybrid ToA/AoA, or multidimensional scaling-based localization \cite{khalil2024isac_mds}. Measurement weights are assigned according to sensing confidence and predicted link reliability, so stable LoS measurements receive higher weights, while uncertain or NLoS-prone measurements are downweighted.

\textcolor{black}{After localization, AIRS-LaaS updates link statistics using localization residuals, outage/beam-failure events, NLoS indicators, and mobility changes. If reports are missing, link/beam failure occurs, model confidence is low, or inference exceeds the latency budget, the controller temporarily switches to a conventional SINR- and geometry-aware selection rule, trading possible overhead/accuracy loss for service continuity. AI-assisted selection resumes once reliable feedback is restored.}

\textcolor{black}{For practical 3GPP/ETSI integration, AIRS-LaaS relies on measurable indicators such as SINR, beam quality, ToA/AoA/Doppler, sensing confidence, localization uncertainty, anchor/beam capability, and AI model status~\cite{3gpp2025aiml}. These indicators support a compact signaling flow that includes capability exposure, periodic or event-triggered ISAC reporting, edge-side anchor--beam activation, and LaaS estimate/confidence/residual feedback. The corresponding functions can be mapped to RRC/NG-RAN capability signaling, RAN/ISAC measurement reports, LPP/NR-positioning-style reports, and OAM-based AI model management for model ID, version, activation, monitoring, and fallback. ETSI ISAC KPIs can further guide sensing-confidence and localization-reliability classes \cite{etsi2025grisc001,huang2025aiml_lcm}.}

\section{Simulation Setup and Performance Evaluation}
\subsection{Simulation Scenario and Parameters}
We evaluate AIRS-LaaS in a 3D $500 \times 500$ m$^2$ urban--suburban UAV-IoT region with four UAV anchors, four terrestrial anchors, and 100 randomly distributed IoT targets, unless otherwise stated. UAV anchors operate at altitudes of 80--150 m, while terrestrial anchors are fixed base stations, roadside units, or IoT gateways. 
\textcolor{black}{To capture UAV ascent/descent and topology dynamics, UAV heights and positions are updated across localization epochs within the service-altitude range. Each update changes the anchor--target distance, LoS probability, geometric contribution, beam-alignment risk, and pilot/beam-refresh overhead; AIRS-LaaS then recomputes anchor--beam scores from the latest sensing and link-quality reports. Fig.~\ref{fig:dynamic_altitude_overhead} shows that positioning overhead increases as UAV altitude variation grows, while AIRS-LaaS limits this growth by activating only high-utility anchor--beam pairs.}

\textcolor{black}{Fig.~\ref{fig:dense_candidate_scalability} shows that localization outage decreases as the candidate anchor--beam set becomes denser. AIRS-LaaS benefits from the larger candidate pool by selecting high-utility links while activating only a budget-limited subset. This confirms that the proposed selection mechanism remains effective under dense UAV-IoT deployments.}

Each anchor uses a directional beam codebook, and each anchor--beam pair is treated as a candidate ISAC link \cite{khalil2024isac_mds}. The carrier frequency is 28 GHz with 100 MHz bandwidth. 
\textcolor{black}{For reproducibility, the 28~GHz link model uses LoS/NLoS path-loss exponents of $2.1/3.5$, log-normal shadowing standard deviations of $3/7$~dB, and an additional NLoS range bias uniformly drawn from $[3,10]$~m. These parameters affect SINR, sensing confidence, and range/AoA reliability used by AIRS-LaaS.} 
Both LoS and NLoS links are considered, where LoS probability depends on anchor--target distance, UAV altitude, and blockage conditions, while NLoS links introduce range bias and angular uncertainty. The localization engine uses hybrid range--angle measurements and downweights unreliable measurements based on sensing-confidence scores. This setting is consistent with recent ISAC and UAV-assisted localization studies \cite{liu2024uav_isac_iot}.

\begin{figure*}[!t]
    \centering

    \subfloat[\protect\textcolor{black}{Localization outage versus error threshold.}]
    {
        \includegraphics[width=0.4\textwidth]
        {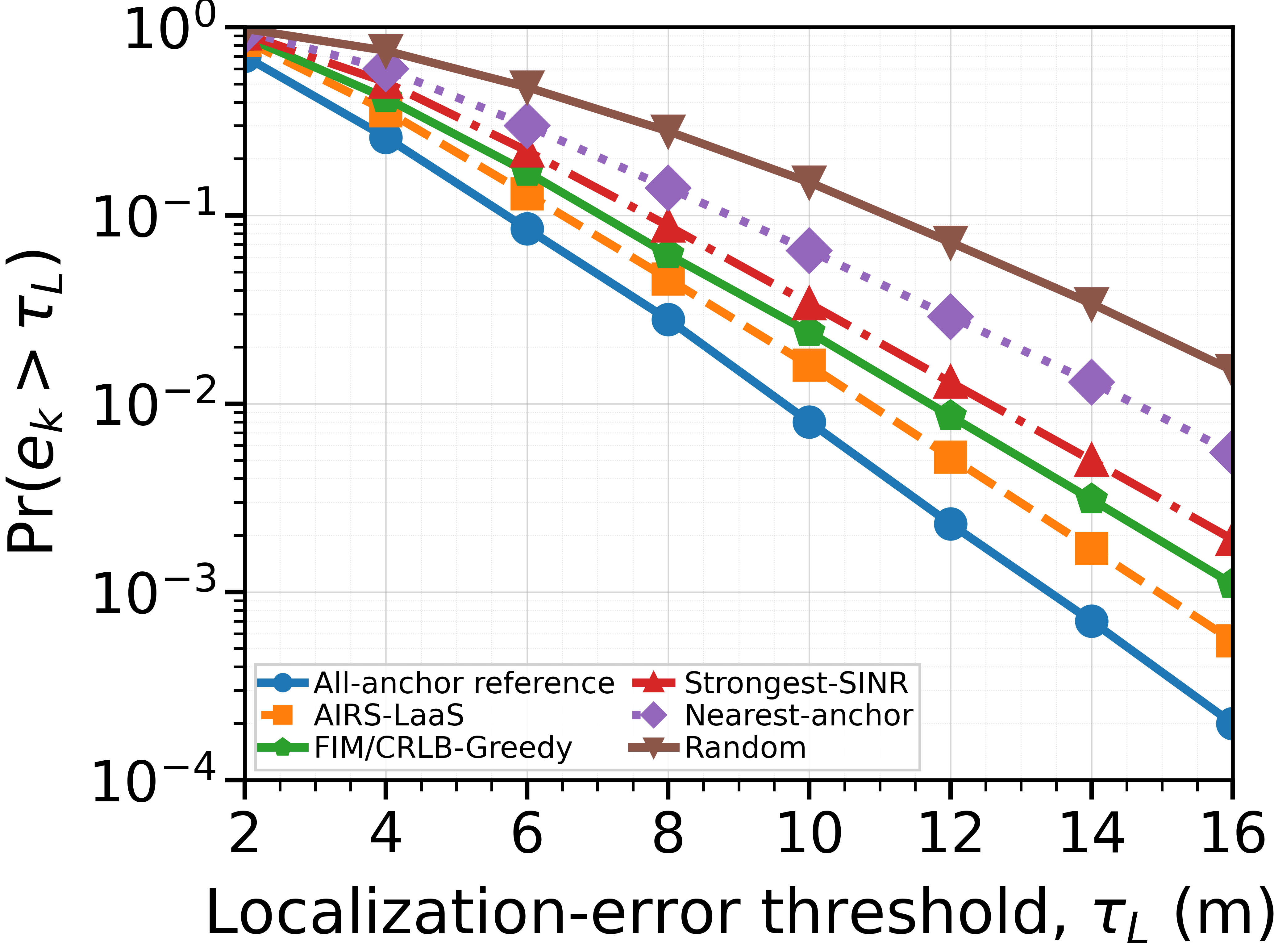}
        \label{fig:loc_outage_threshold}
    }
    \hspace{1.25cm}
    \subfloat[\protect\textcolor{black}{Communication outage versus average SNR.}]
    {
        \includegraphics[width=0.4\textwidth]
        {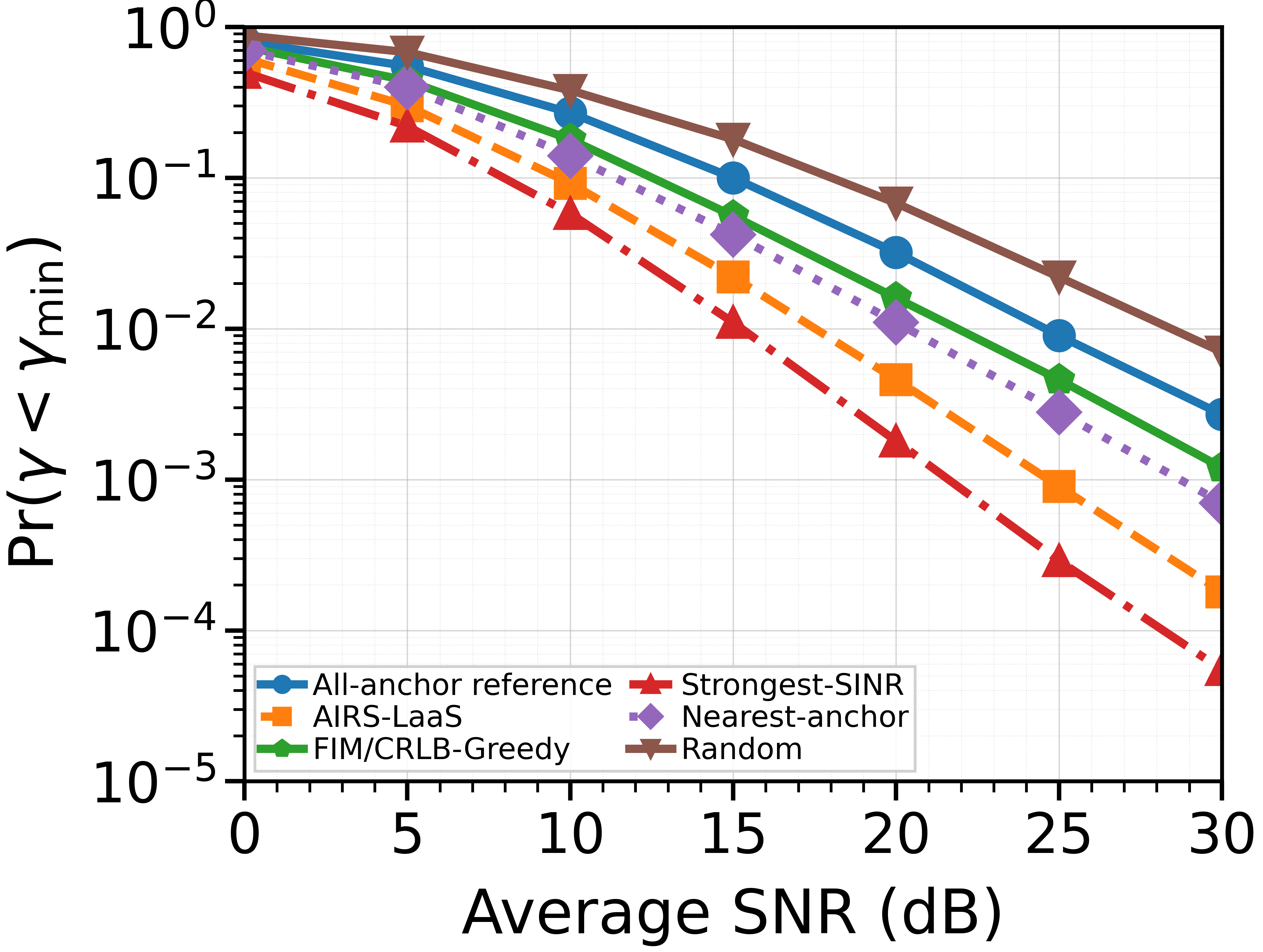}
        \label{fig:comm_outage_snr}
    }

    \vspace{1.5mm}

    \subfloat[\protect\textcolor{black}{Localization outage versus NLoS probability.}]
    {
        \includegraphics[width=0.4\textwidth]
        {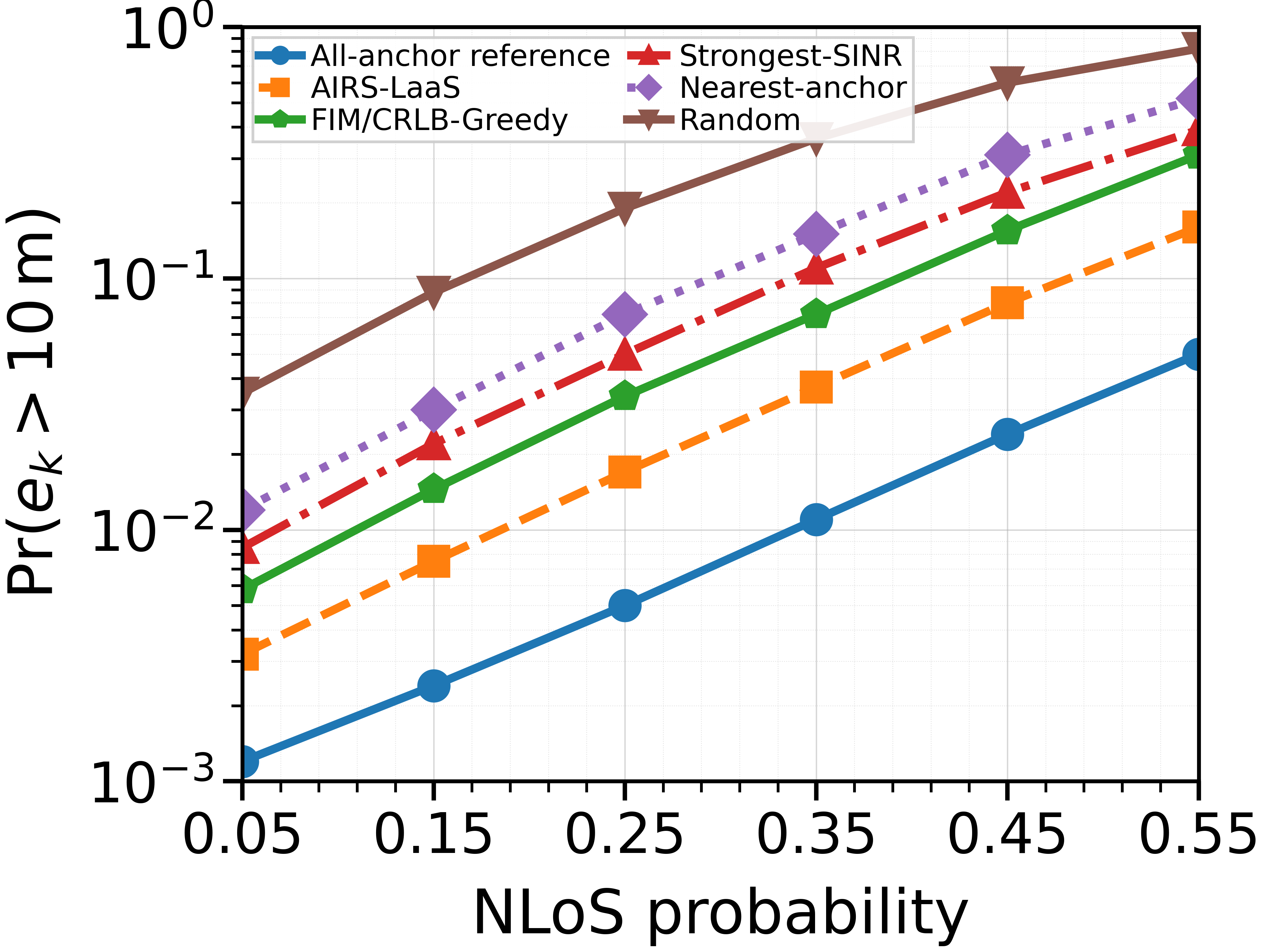}
        \label{fig:loc_outage_nlos}
    }
    \hspace{1.25cm}
    \subfloat[\protect\textcolor{black}{Localization outage versus activated anchor--beam pairs.}]
    {
        \includegraphics[width=0.4\textwidth]
        {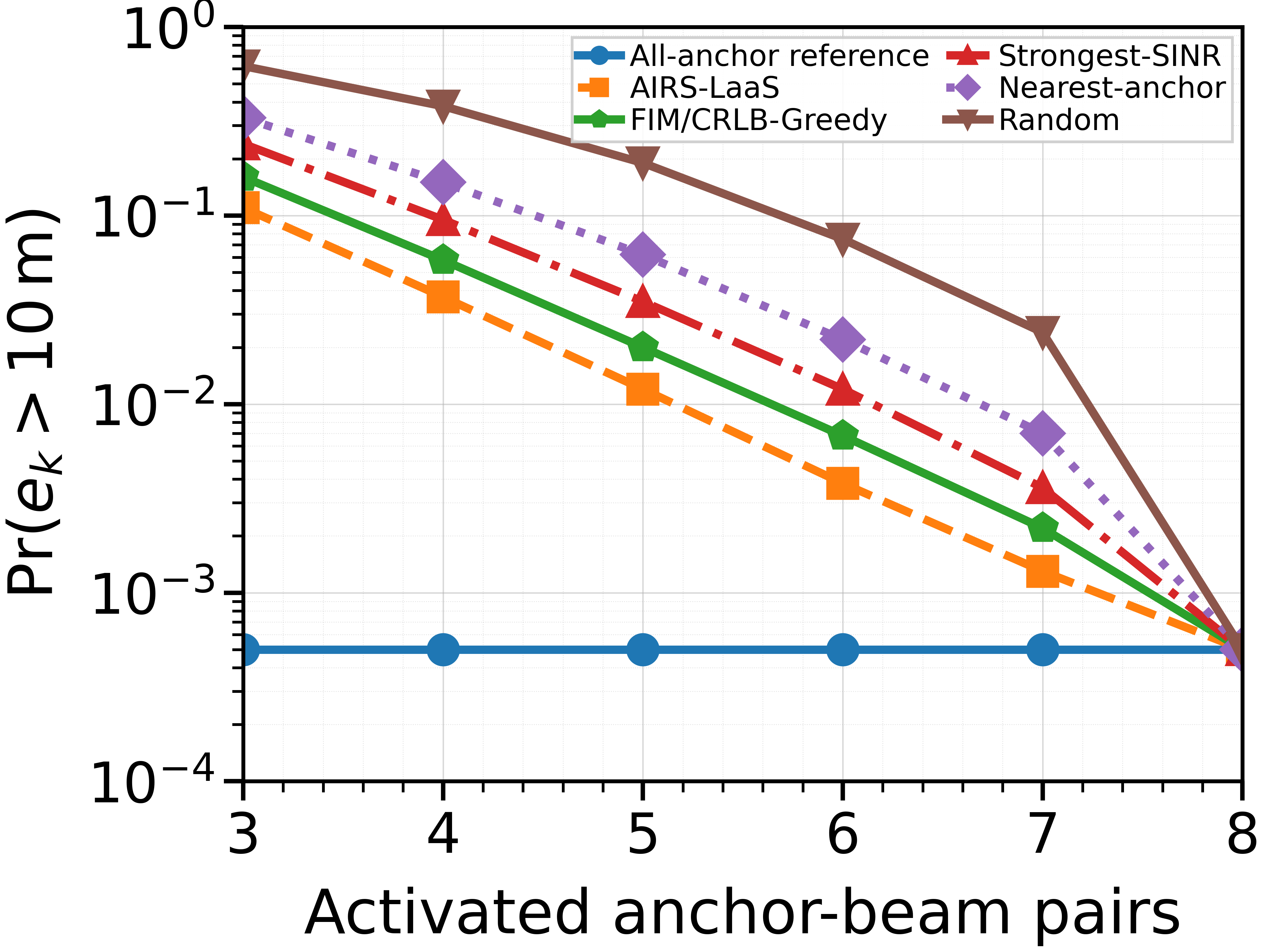}
        \label{fig:loc_outage_active_pairs}
    }

    \caption{\textcolor{black}{AIRS-LaaS performance comparison with benchmark anchor--beam selection schemes.}}
    \label{fig:combined_results}
\end{figure*}

The AI-assisted scoring module uses SINR, sensing confidence, predicted LoS probability, geometric contribution, mobility risk, and resource cost as input features. 
\textcolor{black}{For the reported results, the scoring block is implemented using a gradient-boosting regressor with 100 shallow trees, maximum depth 3, learning rate 0.05, and squared-error loss. Features are min--max normalized, and the model is trained offline using an 80/20 training/validation split from Monte Carlo link observations, with utility labels derived from localization residuals, outage states, and resource cost. The trained model is fixed during online edge inference.} 
In each Monte Carlo trial, the edge controller scores all candidate anchor--beam pairs, selects a subset under the pilot budget, and performs localization using only the selected links.

\subsection{Benchmark Schemes}
AIRS-LaaS is compared with five benchmark schemes. The first is \emph{all-anchor selection}, where all available anchor--beam pairs are activated, providing a high-accuracy but high-overhead reference. The second is \emph{random selection}, in which the same number of anchor--beam pairs as in AIRS-LaaS is selected at random. The third is \emph{nearest-anchor selection}, in which the closest anchors are activated regardless of sensing confidence or beam quality. The fourth is \emph{strongest-SINR selection}, where links with the highest communication quality are selected. 
\textcolor{black}{The fifth benchmark is FIM/CRLB-greedy selection, which sequentially selects the anchor--beam pair with the largest Fisher-information gain, or equivalently the largest reduction in CRLB-oriented localization uncertainty. Unlike AIRS-LaaS, this baseline mainly captures localization geometry and does not jointly account for sensing confidence, LoS probability, mobility risk, communication reliability, and resource cost \cite{khalil2023anchor_activation,khalil2024isac_mds,liu2024uav_isac_iot}.}

For fairness, all benchmark schemes use the same localization estimator after anchor selection. Except for all-anchor selection, all schemes are constrained by the same pilot and beam-training budget, so performance differences mainly arise from the anchor--beam selection strategy.

\subsection{Localization Accuracy and Communication Reliability}
\textcolor{black}{Fig.~\ref{fig:combined_results}(a) shows localization outage probability versus the localization-error threshold $\tau_{\mathrm{L}}$, where outage occurs when $e_k>\tau_{\mathrm{L}}$. The outage probability decreases as the threshold increases. The all-anchor reference provides the lowest outage because it uses the most measurements, while AIRS-LaaS closely follows it with fewer active links.} In contrast, strongest-SINR, nearest-anchor, and random selection suffer higher localization outage because they do not jointly account for sensing confidence, LoS likelihood, and geometric diversity.

\textcolor{black}{Fig.~\ref{fig:combined_results}(b) shows communication outage probability versus average SNR, where outage is defined as $\gamma<\gamma_{\mathrm{min}}$. Strongest-SINR achieves the lowest communication outage because it is designed for link quality.} AIRS-LaaS remains close to this communication-oriented benchmark while also supporting localization reliability. \textcolor{black}{Fig.~\ref{fig:combined_results}(c) shows the localization robustness as the NLoS probability increases. The outage probability increases for all schemes as blockage and biased measurements become more frequent, but AIRS-LaaS remains more robust because it considers LoS probability and sensing confidence before activating anchor--beam pairs. }
\textcolor{black}{These results justify the AI-assisted design: unlike single-factor baselines, AIRS-LaaS jointly uses communication quality, sensing confidence, LoS likelihood, geometric contribution, mobility risk, and resource cost to provide a balanced localization--communication tradeoff. Compared with FIM/CRLB-greedy selection, AIRS-LaaS also accounts for link and sensing reliability, not only geometry.}

\textcolor{black}{To assess the contribution of the utility-score components, we performed a leave-one-feature-out ablation study on AIRS-LaaS. As summarized in Table~\ref{tab:ablation}, removing SINR mainly affects communication reliability; removing sensing confidence/LoS/geometry degrades localization robustness; and removing mobility risk or resource cost affects stability and overhead. This confirms that the full AIRS-LaaS score provides a more balanced localization--communication--overhead tradeoff than partially ablated scoring.}
\begin{table}[!t]
\centering
\caption{\textcolor{black}{Compact ablation summary of AIRS-LaaS.}}
\label{tab:ablation}
\footnotesize
\setlength{\tabcolsep}{2.5pt}
\renewcommand{\arraystretch}{1.03}
\begin{tabular}{p{0.35\columnwidth}p{0.55\columnwidth}}
\toprule
\textcolor{black}{Removed term} & \textcolor{black}{Dominant impact} \\
\midrule
\textcolor{black}{None} & \textcolor{black}{Balanced reliability and overhead} \\
\textcolor{black}{SINR} & \textcolor{black}{Higher communication outage} \\
\textcolor{black}{Sensing conf.} & \textcolor{black}{Less reliable ISAC measurements} \\
\textcolor{black}{LoS probability} & \textcolor{black}{Higher NLoS-induced localization errors} \\
\textcolor{black}{Geometry} & \textcolor{black}{Poorer spatial diversity} \\
\textcolor{black}{Mobility risk} & \textcolor{black}{Less stable dynamic-link selection} \\
\textcolor{black}{Resource cost} & \textcolor{black}{Higher sensing/beam-training overhead} \\
\bottomrule
\end{tabular}
\end{table}

\subsection{Accuracy--Overhead Tradeoff}
\textcolor{black}{Fig.~\ref{fig:combined_results}(d) shows localization outage probability versus the number of activated anchor--beam pairs. Increasing the active set improves reliability by making more measurements available, but it also increases pilot overhead, beam-training delay, energy consumption, and reporting load.} The all-anchor reference achieves the lowest outage but incurs the highest overhead. AIRS-LaaS reaches a low-outage operating region with fewer activated links than full-anchor selection, demonstrating its ability to select high-value links under resource constraints.

\textcolor{black}{The FIM/CRLB-greedy baseline improves over nearest-anchor and random selection by exploiting localization geometry. However, AIRS-LaaS offers a stronger joint trade-off because it also accounts for communication reliability, sensing confidence, LoS likelihood, mobility risk, and resource cost.} These results show that reliable 6G LaaS should not maximize localization accuracy alone; it should jointly balance sensing quality, communication reliability, geometric diversity, mobility awareness, and resource efficiency.

\section{Conclusion}
This article presented AIRS-LaaS, an AI-assisted ISAC Localization-as-a-Service framework for 6G UAV-IoT networks. The proposed framework uses edge intelligence to select a compact set of reliable anchor--beam pairs by jointly considering communication quality, sensing confidence, LoS probability, geometric contribution, mobility risk, and resource cost. By activating only high-utility links before localization, AIRS-LaaS reduces unnecessary pilot signaling, sensing reports, and beam-training overhead while maintaining localization and communication reliability. \textcolor{black}{The evaluation shows that AIRS-LaaS provides a balanced localization--communication--overhead tradeoff compared with all-anchor, FIM/CRLB-greedy, strongest-SINR, nearest-anchor, and random selection schemes under LoS/NLoS and UAV-mobility conditions.} The discussion further highlighted practical deployment aspects, including edge-intelligence KPIs, ISAC measurement reporting, localization confidence, fallback operation, AI model lifecycle management, and privacy-aware data exchange. These findings indicate that AI-assisted ISAC resource selection can support efficient and interoperable 6G UAV-IoT localization services.

\bibliographystyle{IEEEtran}
\bibliography{References}

@techreport{itu2023imt2030,
  author      = {{ITU-R}},
  title       = {{Framework and {O}verall {O}bjectives of the {F}uture {D}evelopment of {IMT} for 2030 and {B}eyond}},
  institution = {International Telecommunication Union, Radiocommunication Sector},
  type        = {Recommendation},
  number      = {ITU-R M.2160-0},
  month       = nov,
  year        = {2023},
  url         = {https://www.itu.int/rec/R-REC-M.2160-0-202311-I/en}
}

@misc{3gpp2025rel19,
  author       = {{3GPP}},
  title        = {{{RAN} {R}el-19 {S}tatus and a {L}ook {B}eyond}},
  howpublished = {{3GPP} {T}echnology {U}pdate},
  month        = apr,
  year         = {2025},
  note         = {By W. Chen, 3GPP TSG RAN Chair},
  url          = {https://www.3gpp.org/technologies/ran-rel-19}
}

@INPROCEEDINGS{heggo2025isac_channel_etsi,
  author={Heggo, Mohammad and Shojaeifard, Arman and Mourad, Alain and Jiang, Chuangxin and Liu, Ruiqi and Liu, Junchen},
  booktitle={IEEE 36th International Symposium on Personal, Indoor and Mobile Radio Communications (PIMRC)}, 
  title={ISAC {C}hannel {M}odels from {ETSI} and {3GPP}}, 
  year={2025},
  volume={},
  number={},
  pages={1-6},
  doi={10.1109/PIMRC62392.2025.11274965}
  }

@techreport{etsi2025grisc001,
  author      = {{ETSI}},
  title       = {{{I}ntegrated {S}ensing and {C}ommunications {(ISAC)}; {U}se {C}ases and {D}eployment {S}cenarios}},
  institution = {European Telecommunications Standards Institute},
  type        = {Group Report},
  number      = {ETSI GR ISC 001 V1.1.1},
  month       = mar,
  year        = {2025},
  url         = {https://www.etsi.org/deliver/etsi_gr/ISC/001_099/001/01.01.01_60/gr_ISC001v010101p.pdf}
}

@misc{3gpp2025aiml,
  author       = {{3GPP}},
  title        = {{Overview of {AI/ML} {R}elated {W}ork in {3GPP}}},
  howpublished = {3GPP News and Events},
  year         = {2025},
  url          = {https://www.3gpp.org/news-events/3gpp-news/ai-ml-2025}
}

@misc{lin2023aiml_nr,
      title={An {O}verview of the {3GPP} {S}tudy on {A}rtificial {I}ntelligence for {5G} {N}ew {R}adio}, 
      author={Xingqin Lin},
      year={2023},
      eprint={2308.05315},
      archivePrefix={arXiv},
      primaryClass={cs.NI},
      url={https://arxiv.org/abs/2308.05315}, 
}

@misc{huang2025aiml_lcm,
      title={{AI/ML} {L}ife {C}ycle {M}anagement for {I}nteroperable {AI} {N}ative {RAN}}, 
      author={Chu-Hsiang Huang and Chao-Kai Wen and Geoffrey Ye Li},
      year={2025},
      eprint={2507.18538},
      archivePrefix={arXiv},
      primaryClass={cs.IT},
      url={https://arxiv.org/abs/2507.18538}, 
}

@article{kaushik2024isac_iot,
  author  = {A. Kaushik and R. Singh and M. Li and H. Luo and S. Dayarathna and R. Senanayake and X. An and R. A. Stirling-Gallacher and W. Shin and M. Di Renzo},
  title   = {{Integrated {S}ensing and {C}ommunications for {IoT}: {S}ynergies with {K}ey {6G} {T}echnology {E}nablers}},
  journal = {IEEE Internet of Things Magazine},
  volume  = {7},
  number  = {5},
  pages   = {136--143},
  month   = sep,
  year    = {2024},
  doi     = {10.1109/IOTM.001.2400052}
}

@ARTICLE{khalili2024uav_isac,
  author={Khalili, Ata and Rezaei, Atefeh and Xu, Dongfang and Dressler, Falko and Schober, Robert},
  journal={IEEE Transactions on Wireless Communications}, 
  title={Efficient {UAV} {H}overing, {R}esource {A}llocation, and {T}rajectory {D}esign for {ISAC} {W}ith {L}imited {B}ackhaul {C}apacity}, 
  year={2024},
  volume={23},
  number={11},
  pages={17635-17650},
  doi={10.1109/TWC.2024.3455370}
  }

@article{khalil2023anchor_activation,
    title = {Energy-{E}fficient {A}nchor {A}ctivation {P}rotocol for {N}on-{C}ooperative {L}ocalization of {I}ndustrial {I}nternet of {T}hings {(IIoT)}},
    author = {Ruhul Amin Khalil and Nasir Saeed and Muhannad Almutiry and Ali Hamdan Alenezi},
    journal = {ICT Express},
    volume = {9},
    number = {5},
    pages = {815-820},
    year = {2023},
    issn = {2405-9595},
    doi = {https://doi.org/10.1016/j.icte.2023.01.004}
}

@article{khalil2024isac_mds,
  author  = {R. A. Khalil and N. Saeed},
  title   = {{Robust {M}ulti-{T}arget {L}ocalization in {ISAC} {S}ystems: {L}everaging {M}ultidimensional {S}caling}},
  journal = {IEEE Open Journal of the Communications Society},
  volume  = {5},
  pages   = {3678--3689},
  year    = {2024},
  doi     = {10.1109/OJCOMS.2024.3413172}
}

@article{liu2024uav_isac_iot,
  author  = {Z. Liu and X. Liu and Y. Liu and V. C. M. Leung and T. S. Durrani},
  title   = {{{UAV} {A}ssisted {I}ntegrated {S}ensing and {C}ommunications for {I}nternet of {T}hings: {3D} {T}rajectory {O}ptimization and {R}esource {A}llocation}},
  journal = {IEEE Transactions on Wireless Communications},
  volume  = {23},
  number  = {8},
  pages   = {8654--8667},
  month   = Aug,
  year    = {2024},
  doi     = {10.1109/TWC.2024.3352985}
}

@INPROCEEDINGS{chen2025generative,
  author={Chen, Jiayuan and Li, Yuxiang and Yi, Changyan and Gong, Shimin},
  booktitle={IEEE INFOCOM 2025 - IEEE Conference on Computer Communications Workshops (INFOCOM WKSHPS)}, 
  title={Generative {AI}-{A}ided {QoE} {M}aximization for {RIS}-{A}ssisted {D}igital {T}win {I}nteraction}, 
  year={2025},
  volume={},
  number={},
  pages={1-6},
  doi={10.1109/INFOCOMWKSHPS65812.2025.11152999}
  }

@ARTICLE{zhu2026data,
  author={Zhu, Fang and Chen, Jiayuan and Wen, Junjie and Yang, Yuye and Yi, Changyan and Tie, Yun and Zhang, Peng and Cai, Jun and Niyato, Dusit and Guizani, Mohsen},
  journal={IEEE Communications Surveys \& Tutorials}, 
  title={From {D}ata {M}irror to {S}mart {C}opilot: {A} {S}urvey on {NextG} {S}emantic {C}ommunication for {P}ropelling {D}igital {T}win {W}orld {I}nto {C}ognitive {S}tage}, 
  year={2026},
  volume={28},
  number={},
  pages={4915-4947},
  doi={10.1109/COMST.2026.3665395}
  }


\end{document}